\documentclass{caosp302}
\usepackage{float}
\usepackage{graphicx}

\articleNo{123}
\pubyear{2007}
\volume{35}
\volnumber{3}
\firstpage{1}
\received{October, 2007}
\accepted{August 28, 2005}

\begin{document}

%
\htitle{Pulsations in HD\,21190 and HD\,218994}
\hauthor{J. F.\,Gonz\'alez, S.\,Hubrig and I.\,Savanov}

\title{A search for pulsational line profile variations\\ in the $\delta$\,Scuti star HD\,21190\\ and the 
Ap Sr star HD\,218994}


%
\author{
       J. F.\,Gonz\'alez\inst{1}
        \and  
        S.\,Hubrig\inst{2}
       \and 
        I.\,Savanov\inst{3}
       }

%
\institute{
           Complejo Astron\'omico El Leoncito, Casilla 467, 5400 San Juan, Argentina \\
           \email{fgonzalez@casleo.gov.ar}
           \and 
          ESO, Casilla 19001, Santiago 19, Chile 
          \and 
           Armagh Observatory, College Hill, Armagh BT61 9DG, NI, UK 
          }

\date{October 8, 2007}

\maketitle

\begin{abstract}
We present the results of our recent search for pulsational line profile variations in high 
time resolution UVES spectra of the most evolved Ap star known, the $\delta$\,Scuti star HD\,21190,
and of the Ap Sr star HD\,218994.
We found that HD\,218994 is an roAp star with a pulsation period of 5.1\,min, 
which makes it the 36$^{th}$ star known to be a roAp star. 
No rapid pulsations have been found in the spectra 
of the $\delta$\,Scuti star HD\,21190. However, we detect moving peaks in the cores of spectral lines, 
which indicate the presence of non-radial pulsations in this star.
\end{abstract}

%
\section{The $\delta$ Scuti star HD\,21190}
This star is a known $\delta$\,Scuti star with a variability period of 3.6\,h, discovered by the 
Hipparcos mission. Koen et al.\ (2001)
reported the spectral type as F2~III SrEuSi, making it the most evolved Ap star known.
Our UVES observations covered  $\sim{}1/4$ of the known $\delta$\,Scuti pulsation period,
showing broad spectral lines ($v \sin i \sim 72$\,km\,s$^{-1}$) with small variable peaks in the 
line profiles (top of Fig.~1).
In the bottom of Fig.~1 we present all observed spectra stacked in two-dimensional images for the same
two spectral regions. It is clearly visible that three peaks are moving evenly towards the red with 
a speed of 21\,km\,s$^{-1}$ per hour. However, no spectral features moving at higher 
frequencies were found in our spectra. To measure the radial velocities of these peaks we 
filtered lower spectral frequencies and used the cross-correlation method.
Similar splitting in line profiles has recently been detected by Yushchenko et al.\ (2005) 
in two other $\delta$ Scuti stars, $\delta$\,Scuti itself and HD\,57749. 

\begin{figure}
\begin{center}
    \includegraphics*[width=0.45\textwidth, bb=130 240 540 552]{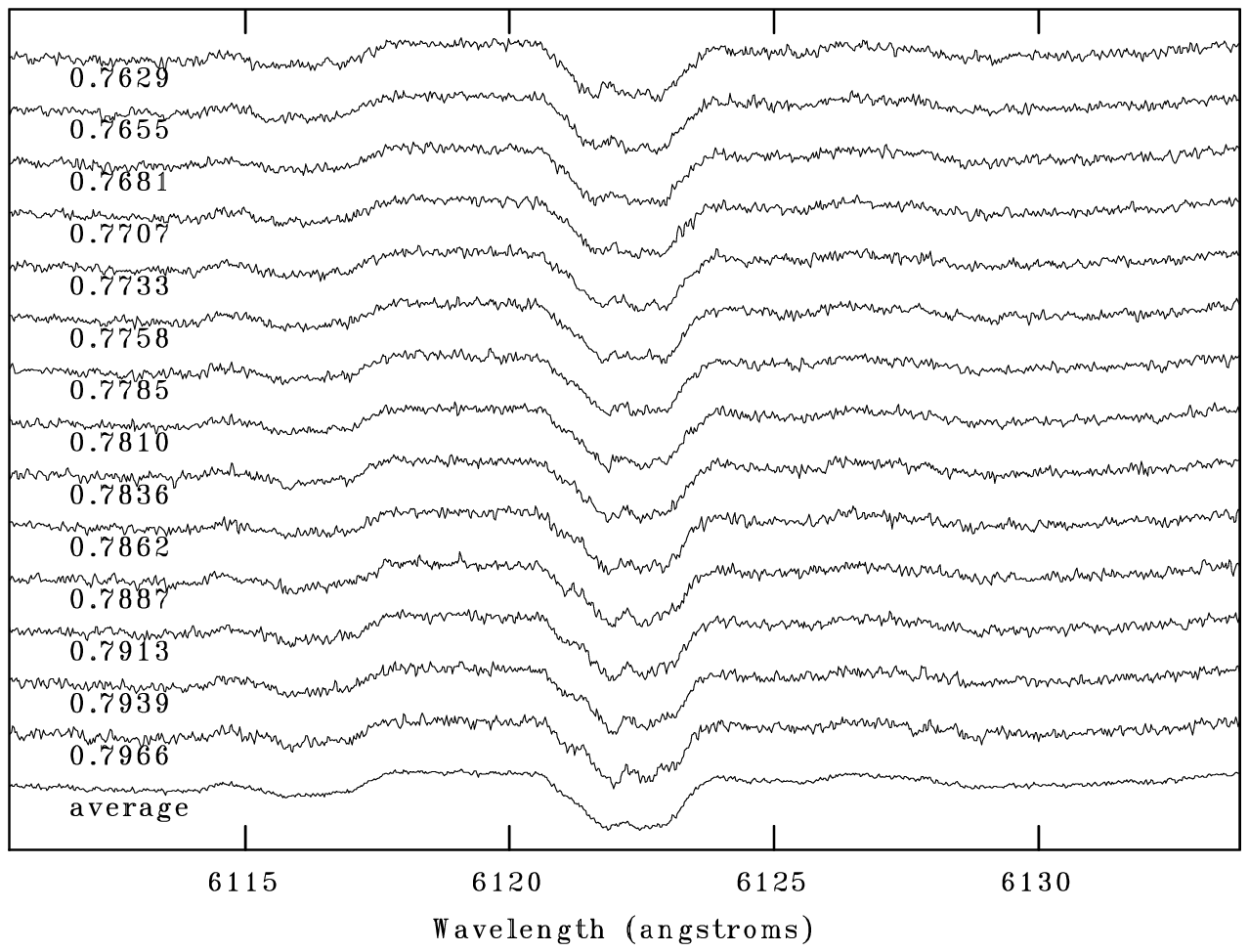}
    \includegraphics*[width=0.45\textwidth,bb=120 240 530 552]{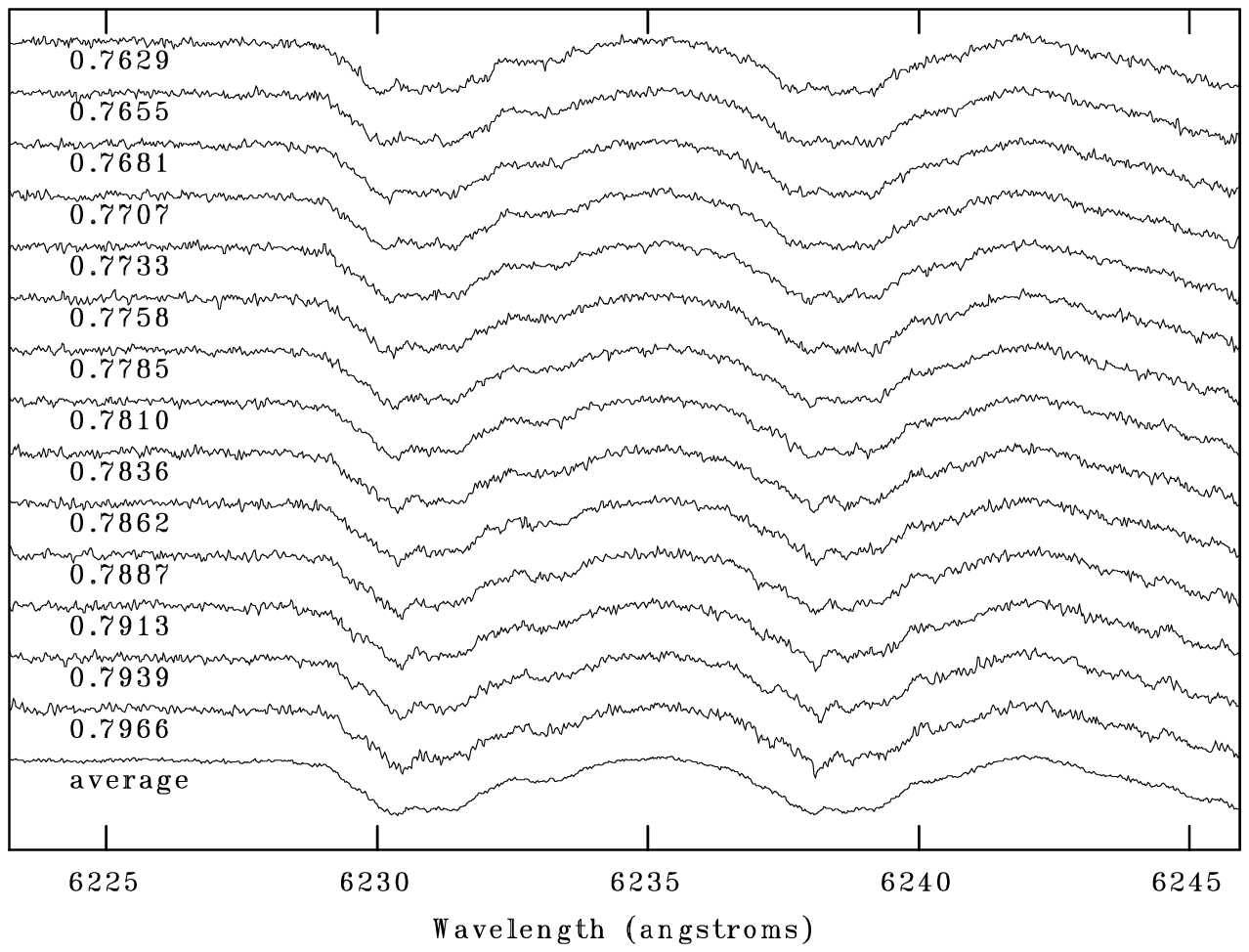}
\includegraphics*[width=0.45\textwidth,height=1.0cm,bb=51 406 601 387]{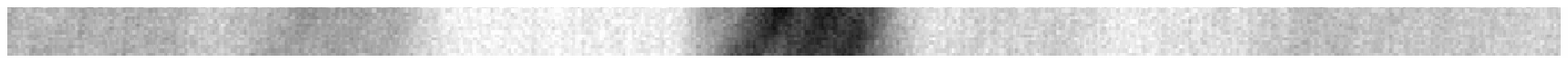}
\includegraphics*[width=0.45\textwidth,height=1.0cm,bb=10 406 617 387]{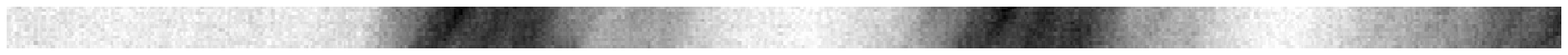}
 \end{center}
\caption{
Spectral profile variability of the Mn\,\textsc{ii} 6122\,\AA{}, Fe\,\textsc{i}
6231\,\AA{}, and Fe\,\textsc{ii} 6238\,\AA{} lines in the UVES spectra of HD\,21190. Spectra are 
labelled with
Julian dates (HJD 2454047). The average spectra are plotted in the bottom of each panel.
The two-dimensional images present, for the same spectral regions, the 14 observed spectra stacked.
Three moving peaks are clearly visible in the cores of all presented spectral lines. 
}
\label{fig:he3}
\end{figure}

\section{The Ap Sr star HD\,218994}
The spectral type in the catalogue of Renson et al.\ (1991) is given as A3\,Sr and from 
Str\"omgren photometry we obtained $T_{\rm  eff} = 7568$\,K  
(Hubrig et al.\ 2000). This Ap star is located in 
the same region of parameter space in which rapid pulsations have 
been detected, but an earlier search for pulsations in the Cape Survey yielded no detection 
(Martinez \& Kurtz 1994). We obtained for this star 15 high time resolution  UVES spectra. From 
these spectra we estimated $v \sin i = 5.2 \pm 0.6$\,km\,s$^{-1}$ using the first zero of the 
Fourier transform of the spectral line profile of the magnetically insensitive lines. 
Our analysis of radial velocity variations of Nd\,\textsc{iii} lines revealed a pulsation 
period of 5.1\,min. 
However, there is a potential ambiguity for the pulsation period 
indicating that also a 14.2\,min period is possible. Longer time series 
with higher temporal resolution are needed for a more 
detailed analysis of the pulsational frequencies.
From fitting 
the radial velocity variations of the Nd\,\textsc{iii} line at 
$\lambda$\,6327 with a sinusoidal curve we obtained 
a semi-amplitude of 0.51\,km\,s$^{-1}$.


{}
\end{document}